\documentclass[rnote]{aa} % for the research notes
\usepackage{graphicx}
\usepackage{natbib}
\usepackage{txfonts}
\begin{document}

   \title{
Spectroscopic identification of a redshift 1.55 supernova host galaxy from the Subaru Deep Field Supernova Survey\thanks{Based on observations made with ESO telescopes at the La Silla Paranal Observatory under program ID 089.A-0739}
   }

   \author{Teddy F. Frederiksen\inst{1} \and Or Graur\inst{2}\fnmsep\inst{3} \and
Jens Hjorth\inst{1} \and Dan Maoz\inst{2} \and Dovi Poznanski\inst{2}
}

   \institute{
Dark Cosmology Centre, Niels Bohr Institute, University of Copenhagen, Juliane Maries Vej 30, 2100 Copenhagen, Denmark
\email{teddy@dark-cosmology.dk}
\and
School of Physics and Astronomy, Tel-Aviv University, Tel-Aviv 69978, Israel
\and
Department of Astrophysics, American Museum of Natural History, Central Park West and 79th Street, New York, NY 10024-5192, USA
}

\authorrunning{Frederiksen et al.}
\titlerunning{Spectroscopically confirmed SDF SN host}

%   \date{Received September 15, 1996; accepted March 16, 1997}

% \abstract{}{}{}{}{} 
% 5 {} token are mandatory
 
  \abstract
   {The Subaru Deep Field (SDF) Supernova Survey discovered ten Type Ia supernovae (SNe Ia) in the redshift range $1.5<z<2.0$, determined solely from photometric redshifts of the host galaxies. However, photometric redshifts might be biased, and the SN sample could be contaminated by active galactic nuclei (AGNs).}
   {We aim to obtain the first robust redshift measurement and classification of a $z > 1.5$ SDF SN Ia host galaxy candidate}
   {We use the X-shooter (U-to-K-band) spectrograph on the Very Large Telescope {to allow the detection of different emission lines in a wide spectral range}.}
   {We measure a spectroscopic redshift of $1.54563 \pm 0.00027$ of hSDF0705.25, consistent with its photometric redshift of $1.552 \pm 0.018$. From the strong emission-line spectrum we rule out AGN activity, thereby confirming the optical transient as a SN. The host galaxy follows the fundamental metallicity relation showing that the properties of this high-redshift SN Ia host galaxy is similar to other field galaxies.}
   {Spectroscopic confirmation of additional SDF SN hosts would be required to confirm the cosmic SN rate evolution measured in the SDF.}


   \maketitle
%
%________________________________________________________________

\section{INTRODUCTION}

The nature of the progenitor stellar systems of Type Ia supernovae (SNe Ia) remains a mystery \cite[see][for reviews]{Howell2011,Maoz2011review,2013arXiv1312.0628M}.
While both circumstantial and direct lines of evidence point to a carbon-oxygen white dwarf \citep[WD;][]{Nugent2011,Bloom2012} as the progenitor, the otherwise stable WD must be ignited \citep[see][for a review]{Leibundgut2000}.
The current consensus is that the carbon in the core of the WD is ignited as a result of the buildup of pressure, or temperature, resulting from mass accretion from a companion star in a binary system.
The two leading scenarios for the nature of the progenitor binary system are the single degenerate scenario \citep[SD;][]{Whelan1973,1982ApJ...253..798N}, which contends that the WD accretes mass from a main-sequence, helium, or giant star; and the double degenerate scenario \citep[DD;][]{Iben1984,Webbink1984}, in which the WD merges with a second CO~WD through loss of angular momentum and energy to gravitational waves.

{Volumetric SN rates provide strong constraints on SN progenitor models. An important discriminator of explosion scenarios is the delay time distribution (DTD), which quantifies the distribution of times from progenitor formation to explosion.}

%{The SNe were used to measure rates at these redshifts.}
A SN survey in the Subaru Deep Field (SDF) was conducted with the Subaru 8.2 m Telescope. \citet[G11]{Graur2011} discovered 150 candidate SNe, of which 28 (10) were  classified as $1<z<1.5$ ($z>1.5$) SNe~Ia.
Each SN candidate in the SDF was observed in one of four independent epochs in the {\textit{R}}, {\textit{i$^\prime$}}, and {\textit{z$^\prime$}}\ bands.
Consequently, variable active galactic nuclei (AGNs) can be mistaken for SNe.
\citet[G11]{Graur2011} identified interloping AGNs using a catalog of known variable AGNs in the SDF (provided by T. Morokuma) and by culling SN candidates that appeared in more than one of the four survey epochs (see their Sect. 3.1 for a detailed description).
However, without the use of spectroscopy, the possibility that some of the $z>1.5$ SDF SNe~Ia are in fact AGNs could not be ruled out.

Combining {the measured SN rates} with SN rates at other redshifts, and comparing to different realizations of the cosmic star-formation history, G11 set constraints on the SN Ia DTD, which in turn impacts on the progenitor question.
However, the classification of the SNe discovered in the SDF is purely photometric and depends on the redshift of the host galaxy.
The redshifts of most of the SDF SN host galaxies, including those at $z>1.5$, are photometric redshifts (photo-$z$'s).
These photo-$z$'s are based on photometry in 11 bands, from the far-ultraviolet (UV) to the near-infrared (IR), trained on hundreds of galaxies in the field with spectroscopy.
Yet, because of the inherent difficulty in obtaining spectroscopic redshifts for early-type galaxies because of the lack of strong emission lines, training the photo-$z$ method used by G11 at high redshift is difficult.
There could be systematic biases in the redshift estimates in that range, biases that are not accounted for in the formal uncertainty.
Because of the small number of SN candidates, even a few catastrophic photo-$z$ failures or contamination by unidentified AGNs could strongly distort the inferred DTD.
In order to determine whether the G11 $z>1.5$ rate suffers from such systematic biases, the host galaxies of the SN candidates must be observed spectroscopically.

\begin{figure*}
   \resizebox{\hsize}{!}
            {\includegraphics{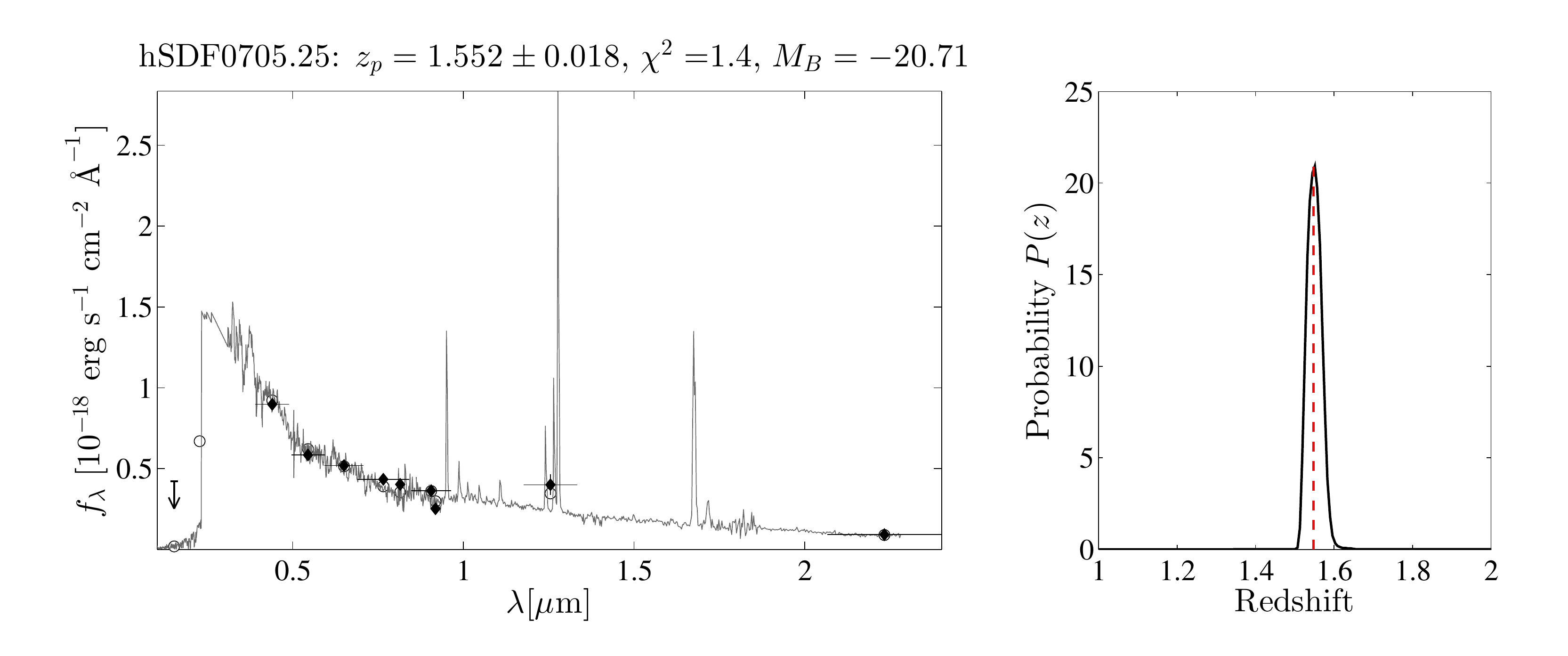}}
	\caption{{\scriptsize ZEBRA} fit and resultant $z$-PDF of hSDF0705.25. The left panel shows the actual photometry (filled circles), the best-fitting galaxy template (solid line), and its synthetic photometry (empty circles). The vertical error bars denote the photometric uncertainty, and the horizontal error bars show the width of the filter. The $1\sigma$ upper limit on the photometry in the {\it GALEX FUV} band is shown as the downturned arrow. The header gives the designation of the SN host galaxy, most probable photo-$z$ ($z_p$), the $\chi^2$ per degree of freedom of the fit, and the absolute {\textit{B}}-band magnitude the galaxy would have at $z_p$. The right panel shows the resultant $z$-PDF peaking at $z_p=1.552\pm0.018$. The red line marks the spectroscopic redshift obtained in this work.
	}
	\label{fig_sed}
\end{figure*}

The {\textit{i$^\prime$}} = 24 mag host galaxy of SNSDF0705.25, denoted hSDF\-0705.25, was typed as a Sbc galaxy with a sharp redshift probability distribution function ($z$-PDF) that peaked at $z_p = 1.552 \pm 0.018$.
In Figure~\ref{fig_sed} we show the photometry of hSDF0705.25, along with the best-fitting galaxy spectral-energy distribution and resultant $z$-PDF.

In this research note we present a VLT/X-shooter emission-line spectrum, derive the spectroscopic redshift and classify AGN vs.\ star-formation activity of the SN host galaxy hSDF0705.25. We also place constraints on the metallicity and star-formation rate of the host galaxy. From broad-band photometry we determine the stellar mass to place the host on the fundamental metallicity relation \citep[FMR;][]{2010MNRAS.408.2115M,2011MNRAS.414.1263M}.
Throughout this research note we assume a flat $\Lambda$CDM cosmology with $H_0=70$~km~s$^{-1}$ and $\Omega_m=0.3$.

\section{DATA}\label{sec_obs}

\begin{figure*}
	\resizebox{\hsize}{!}{\includegraphics{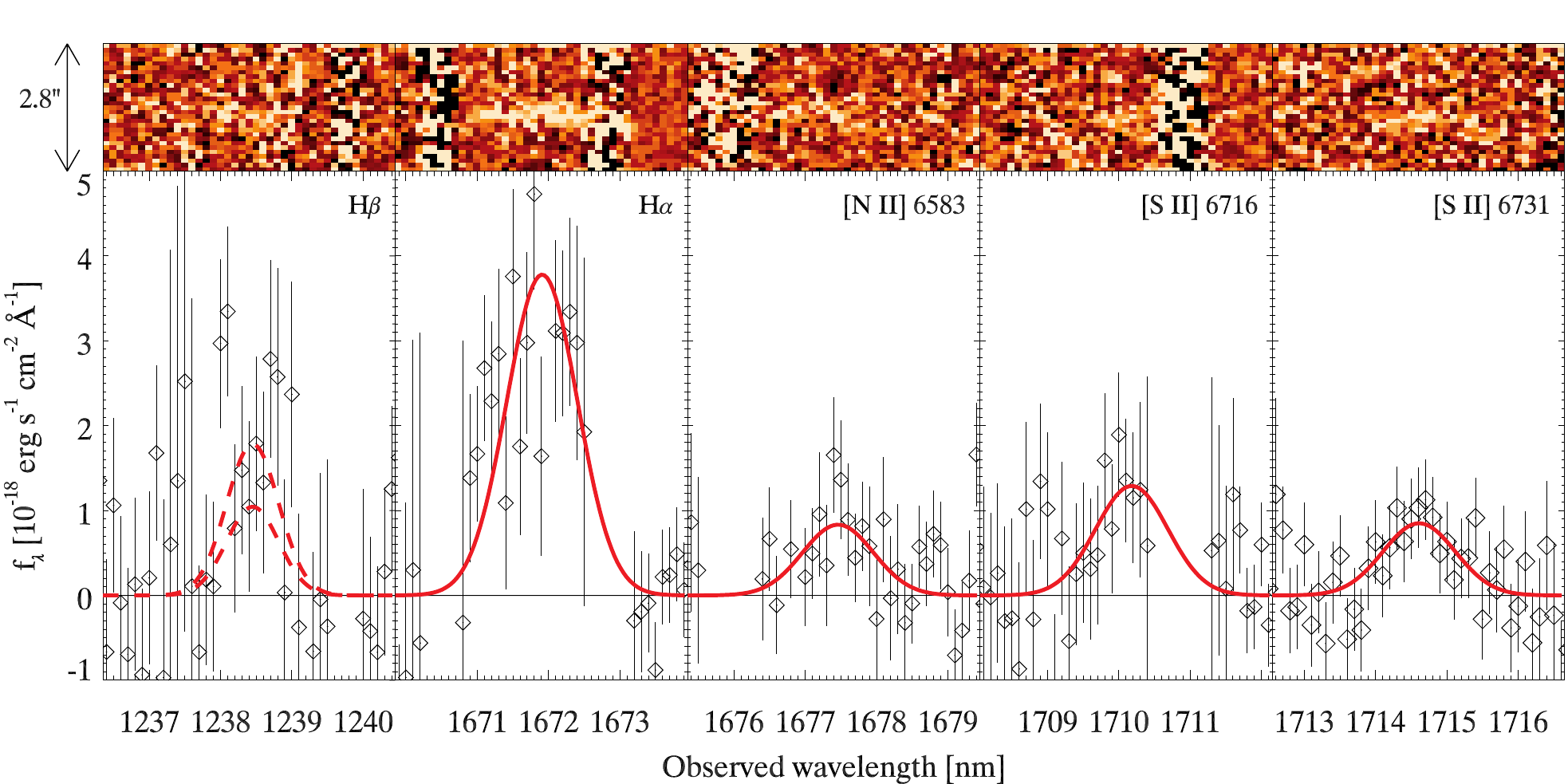}}
	\caption{X-shooter 1D and 2D spectrum of hSDF0705.25. The spectrum has been corrected for Galactic extinction. The solid (red) line shows the fit to the emission lines assuming a Gaussian line profile. The dashed line shows a scaled version of the H$\alpha$ line assuming the same FWHM (in velocity units) and extinction in the host of $E(B-V)=0$ mag (upper) and $E(B-V)=0.5$ mag (lower). Color version available online.
	\label{fig_spec}}
\end{figure*}
The spectrum of hSDF0705.25  was obtained on 2012 April 23 with the X-shooter spectrograph \citep{2006SPIE.6269E..98D,2011A&A...536A.105V} at the Very Large Telescope (VLT) at Cerro Paranal, Chile.
We used an ABBA on-source nodding template with an exposure time of 1.3 hr (4$\times$1200 sec) and a $0\farcs9$ slit.\footnote{A $1\farcs0$ slit in the UVB arm.} The spectrum was obtained under clear conditions. For details on high-redshift emission-line spectroscopy with X-shooter, see \citet{Frederiksen12}.

The X-shooter spectra were reduced using the official X-shooter pipeline\footnote{See, \url{http://www.eso.org/sci/software/pipelines/}} v1.3.7.
The extraction of the object spectrum was conducted with our own IDL script, and flux calibration was done using the flux standard star, LTT~3218.\footnote{See \url{http://www.eso.org/sci/observing/tools/standards/spectra/ltt3218.html}}

The broad-band photometry was taken from G11.

\section{ANALYSIS}\label{sec_analysis}
The flux-calibrated spectrum is corrected for Galactic extinction and slit loss. The Galactic extinction along the line of sight to hSDF0705.25 is $A_V=0.042$ mag\footnote{Quoted from the NASA/IPAC Extragalactic Database (NED) website: \url{http://ned.ipac.caltech.edu/}} \citep{2011ApJ...737..103S} and we assume the Galactic extinction law of \cite{1999PASP..111...63F}. The slit loss correction assumes a $\lambda^{-0.2}$ variation in seeing.\footnote{The standard star itself was observed with a 5\arcsec{} slit, making slit losses negligible.}

In the available atmospheric transmission windows, we detect a strong emission line at $1671.88$~nm which we identify as H$\alpha$ at $z=1.54563\pm0.00027$. This identification is supported by the detection of H$\beta$, [\ion{N}{ii}]~6583, and [\ion{S}{ii}]~6716,6731 at the same redshift.
We do not detect [\ion{O}{iii}]~4959,5007 in the spectrum {as it is obscured by a broad complex of sky lines}.
We also detect [\ion{O}{ii}]~3726,3729 in the VIS arm of X-shooter, but the blue part of the doublet is obscured by a skyline. Without the detection of [\ion{O}{iii}]~5007 we cannot derive a metallicity from the $R_{23}$-ratio. We therefore only consider the emission lines present in the NIR arm as presented in Figure \ref{fig_spec}.

The detected emission lines in the spectrum are fitted with {a set of Gaussian line profiles with a common FWHM (in velocity units). The centroid and normalization of each line is left free to vary.} The flux is calculated from the fit.
The H$\beta$ line is severely affected by noise and therefore not fitted.
The H$\alpha$ line seems to have a larger line width then other lines. which might be an artifact due to the masked skylines on either side of the line. The masked skylines allow for a wider fit and the central part of the line does not  constrain the fit to a more narrow profile.
For H$\beta$, we plot in Figure~\ref{fig_spec} a scaled-down version of the H$\alpha$ line assuming case-B recombination, central wavelength and FWHM (in velocity units) as for H$\alpha$, and intrinsic reddening in the host of $E(B-V)$ of zero or 0.5 mag \citep{2001PASP..113.1449C} to illustrate the variation allowed by the spectrum.

The emission line ratios $N2=\log([\ion{N}{ii}] \, 6583 / H\alpha)$ and $O3=\log([\ion{O}{iii}] \, 5007 / H\beta)$ are among the main diagnostics for discriminating between star formation and AGN activity by way of the \citet*[BPT]{1981PASP...93....5B} diagram.
The $N2$ diagnostic places the host in the star forming region of the BPT diagram.
%This is also supported by the $O3$ diagnostic, which provides an upper limit.
%The upper limit on $O3$ is calculated using the $3\sigma$ upper limit on [\ion{O}{iii}] and the predicted H$\beta$ flux from the H$\alpha$ flux. We calculate the H$\beta$ flux assuming $E(B-V)=0$ or 0.5 mag to make sure the classification does not depend on the assumed amount of extinction in the host galaxy.
%In both cases the host galaxy falls within the star-forming region of the BPT diagram. {The $N2$ value is enough to place the host in the star forming part of the BPT diagram regardless of the more uncertain value of the $O3$-diagnostic.}
%\includegraphics[height=2.5ex]{BPT}
%\footnote{This is an in-text representation of the BPT plot, showing the host galaxy in the star-forming region. The main point is to show the host as star forming, AGN, or composite, while not using a full figure on something that only contains a few bits of information. C.f. the concept of ``Sparklines'' used in economy.}
%\citep[$N2=-0.88$, $O3<0.28$, see e.g.][their Figure 5]{Frederiksen12}.
The lack of a broad-line component in the emission lines (see Table \ref{tab_flux}) and the absence of strong emission from C~IV~1550 and Mg~II~2799 also supports the conclusion that the line-flux is dominated by star formation and not AGN activity.

{From the $N2$ line ratio we derive a metallicity of $8.53\pm0.08$ \citep[PP04. {The intrinsic scatter of this calibration is $\sim$0.2~dex}]{2004MNRAS.348L..59P}.
{In comparison, the calibration of \citet[M08]{2008A&A...488..463M} gives a metallicity of $8.79\pm0.12$.}
% We use the emission-line calibrations of \citet[KD02]{2002ApJS..142...35K} to constrain the metallicity. (In addition we get a constraint on the ionization parameter $q$.)}
%We make use of the emission-line ratios $N2$ and $N2O3=\log([\ion{N}{ii}] \, 6583/[\ion{O}{iii}] \,5007)$. The $N2$ ratio {(like the PP04 metallicity)} places the host galaxy in the low-metallicity regime at $12+\log(O/H)<8.55$. The $N2O3$ ratio gives us an upper limit of $12+\log(O/H)<8.0$ (and ionization parameter $q<2 \times 10^{7}$ cm s$^{-1}$).

We convert the rest-frame H$\alpha$ luminosity into a star-formation rate (SFR) using the calibration of \cite{1998ARA&A..36..189K}, rescaled to a \citet{2003PASP..115..763C} initial mass function (IMF). The observed SFR, $4.0\pm0.7$~M$_\odot$~yr$^{-1}$,  represents a lower limit to the intrinsic SFR of the host galaxy, due to the unconstrained extinction in the host galaxy.

The photo-$z$ fitter {\scriptsize ZEBRA} \citep{Feldmann2006} is not suited for deriving stellar parameters like the mass of the galaxy. We therefore fit the Subaru+UKIRT ({\textit{B}}, {\textit{V}}, {\textit{R}}, {\textit{i$^\prime$}}, {\textit{z$^\prime$}}, \textit{J} and \textit{K}) photometric measurements (corrected for foreground extinction in the same way as the spectrum) using the {\scriptsize FAST} SED fitter \citep{2009ApJ...700..221K}. We derive the intrinsic extinction, stellar mass, and stellar age of the host galaxy (see Table \ref{tab_results}). The extinction in the host is not very well constrained ($0<A_V<1.1$) so we assume $A_V=0$ in our further analysis. We calculate the specific SFR (sSFR, see Table 2) using the H$\alpha$ SFR and the stellar mass from {\scriptsize FAST}.

\section{DISCUSSION}
\begin{table}
\caption{Spectroscopic summary}\label{tab_flux}
\centering
\begin{tabular}{l r r}
\hline\hline
		\multicolumn{1}{c}{Line} &
		\multicolumn{1}{c}{Wavelength} &
		%\multicolumn{1}{c}{Obs. FWHM$^{a}$} &
		%\multicolumn{1}{c}{Int. FWHM$^{a}$} &
		\multicolumn{1}{c}{Flux} \\
		&
		\multicolumn{1}{c}{(nm)} &
		%\multicolumn{1}{c}{(km s$^{-1}$)} &
		%\multicolumn{1}{c}{(km s$^{-1}$)} &
		\multicolumn{1}{c}{($10^{-19}$ erg s$^{-1}$ cm$^{-2}$)} \\
\hline
		H$\alpha$ & $1671.91\pm0.05$
		%& $88.9\pm7.6$ & $68.2\pm7.6$ 
		& $37.8\pm4.3$ \\
		
		[\ion{N}{ii}] 6583 & $1677.46\pm0.12$ 
		%& $88.9\pm7.6$ & $68.2\pm7.6$ 
		& $\;8.3\pm2.2$ \\

		[\ion{S}{ii}] 6716 & $1711.18\pm0.13$ 
		%& $88.9\pm7.6$ & $68.2\pm7.6$ 
		& $12.9\pm2.9$ \\
		
		[\ion{S}{ii}] 6731 & $1714.64\pm0.09$ 
		%& $88.9\pm7.6$ & $68.2\pm7.6$ 
		& $\;8.5\pm1.7$ \\
		\hline
		[\ion{O}{iii}] 5007$^{b}$ & \multicolumn{1}{c}{\ldots} 
		%& \multicolumn{1}{c}{\ldots} & \multicolumn{1}{c}{\ldots} 
		& $<23.2\pm10.4$ \\
		
		[\ion{O}{iii}] 4959$^{b}$ & \multicolumn{1}{c}{\ldots} 
		%& \multicolumn{1}{c}{\ldots} & \multicolumn{1}{c}{\ldots} 
		& $< \ \ 9.1\pm10.4$ \\
		\hline\hline
	\end{tabular}
	\tablefoot{
%		${a)}$ {{All lines were fitted at the same time with a common FWHM(observed)${}=88.9\pm7.6$~km~s$^{-1}$ corresponding to FWHM(intrinsic)${}=68.2\pm7.6$~km~s$^{-1}$}. The instrumental resolution is 57~km~s$^{-1}$.}
		${a)}$ {This is the 3$\sigma$ upper limit on the flux, {measured from the variance in the spectrum at the location of the lines.}}
	}
\end{table}%
\begin{table}
\caption{Derived properties of the SN host galaxy hSDF0705.25}\label{tab_results}
\centering
\begin{tabular}{l r r}
	\hline\hline
	\multicolumn{1}{c}{Property} &
	\multicolumn{1}{c}{Value} \\
	\hline
		Redshift & $z=1.54768\pm0.00008$ \\
		Star-formation rate$^{ab}$ & $SFR=3.2\pm0.5$ M$_\odot$ yr$^{-1}$ \\
		Metallicity (PP04) & $12+\log(\rm O/\rm H) = 8.5\pm0.2$ \\
		Metallicity (M08) & $12+\log(\rm O/\rm H) = 8.8\pm0.2$ \\
%		Ionization parameter (KD02) & $q<4 \times 10^7$ cm s$^{-1}$ \\
		\hline
		\smallskip
		Stellar mass$^{c}$ & $\log(M_* [\rm M_\odot])= 9.46^{+0.23}_{-0.08}$ \\
		\smallskip
		Specific SFR$^{d}$ & $\log({\rm sSFR} [\rm{yr}^{-1}])=-8.96^{+0.10}_{-0.24}$ \\
		Host extinction$^{c}$ & $A_{V,\rm{host}}=0.4^{+0.7}_{-0.4}$ mag \\
		Stellar age$^{c}$ & $\log(t_* [\rm{yr}])=7.9^{+0.3}_{-0.6}$ \\
	\hline\hline
	\end{tabular}
	\tablefoot{
		${a)}$ {Assuming a \cite{2003PASP..115..763C} IMF.}
		${b)}$ {SFR is a lower limit as extinction in the host galaxy would make the intrinsic SFR higher {by $34\%$ ($A_V=0.4$)}.}
		${c)}$ {Using an exponentially declining star-formation history, redshift fixed to the spectroscopic redshift, and metallicity fixed to $Z=0.008$ (i.e., $12+\log(\rm O/\rm H) \sim 8.4$).}
		${d)}$ {Assuming $A_V=0.4$ corresponds to a shift of $\Delta\log({\rm sSFR})=+0.127$.}
	}
\end{table}%

We find a highly star-forming (i.e., high sSFR), low-metallicity SN host galaxy \cite[see][for the discussion of another high SF low metallicity SN host at similar redshift]{Frederiksen12}. The derived SFR may be affected by extinction in the host galaxy, but because of the low signal-to-noise ratio of the H$\beta$ line we are not able to place any strong constraints on it. Likewise, the extinction derived from SED fitting does not provide a strong constraint. {If we use the best fit SED value of $A_V=0.4$~mag the SFR would increase by 34\%.}

The stellar mass and SFR of hSDF0705.25 places it midway between the $z=1$ and $z=2$ main-sequence of star-forming galaxies defined in \citet{2007ApJ...670..156D} and \citet{2007A&A...468...33E}. As hSDF0705.25 falls on the main sequence at its redshift, it can be classified as an average star-forming galaxy at its redshift.
Alternatively, the sSFR can be used to define whether a galaxy is passive, star forming or a starburst galaxy. Using the definition of \citet[see their Figure 6]{2006ApJ...648..868S}, the high sSFR of hSDF0705.25 makes it  a starburst galaxy.
Such galaxies are representative of the ``prompt'' population of the DTD of SNe Ia \citep{2005A&A...433..807M,2005ApJ...629L..85S,Maoz2011review}.

%Having two different measures of metallicity that do not agree is not unusual, as the absolute value of the metallicity depends on the calibration (also called the metallicity scale) used. \cite{2008ApJ...681.1183K} made the comparison between different metallicity calibrations and determined transformations to translate from one metallicity scale to another. From the transformation we expect PP04 to be systematically lower then KD02. The metallicity derived for this host galaxy is inconsistent with this tendency as the upper limit on the KD02 metallicity is $\sim2\sigma$ lower then the PP04 metallicity. This is can be because of extinction effects or to some differences between theoretical and empirical calibrations that are not captured in KE04. The extinction effect is most important for the $N2O3$ line ratio (because of the wide separation between the lines). We have tried to account for extinction by assuming a conservative upper limit on $N2O3$ (assuming $E(B-V)=0.25$). We therefore believe that the difference is because of differences between theoretical and empirical calibrations.

\begin{figure}
	\centering
	\includegraphics[width=0.5\textwidth]{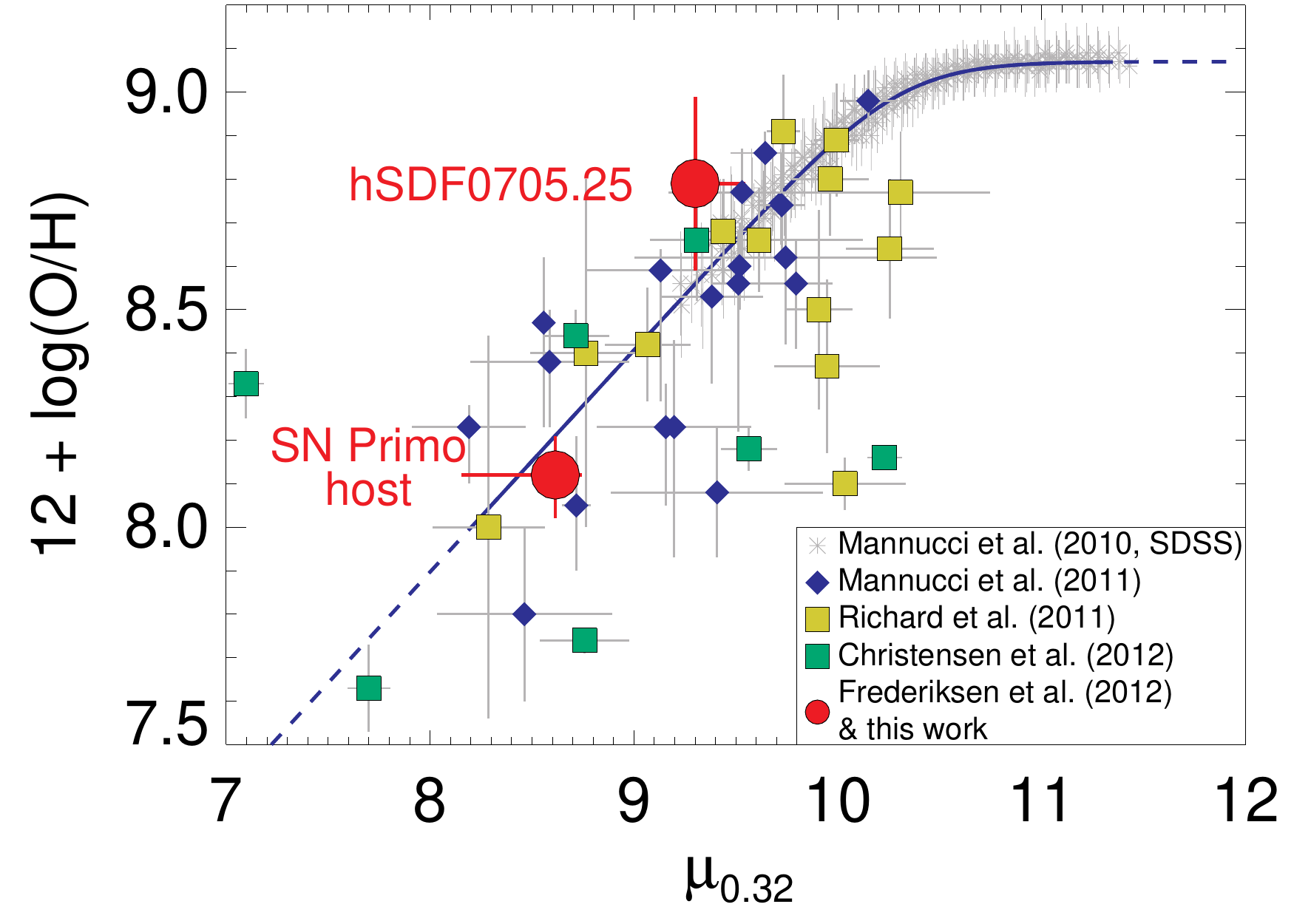}
	\caption{The Fundamental Metallicity Relation (FMR). The parameter $\mu_{0.32}=\log(M_*)-0.32\log(SFR)$ is the projection in the stellar mass, SFR plane.
	The two circles (red) are the two high-redshift SN Ia host galaxies with measured gas-phase metallicity, hSDF0705.25 and SN Primo \citep{Frederiksen12}. The crosses are the binned data from the Sloan Digital Sky Survey \citep[SDSS]{2000AJ....120.1579Y} used in \citet{2010MNRAS.408.2115M}. The diamonds (blue) are GRB host galaxies used to calibrate the low metallicity slope in \citet[$z<1$]{2011MNRAS.414.1263M}. The squares are lensed galaxies from \citet[yellow, $z>1$]{2012MNRAS.427.1953C} and \citet[green, $z>1.5$]{2011MNRAS.413..643R}.
	The dark (blue) line is the parameterization of \citet{2011MNRAS.414.1263M} (solid in the range of validity, extrapolation in dashed blue, high metallicity as constant, low metallicity as straight line).
	The two SN Ia host galaxies ($z=1.5$) are fully consistent with the FMR of star-forming field galaxies.
	\label{fig_fmr}}
\end{figure}
The combination of SFR, stellar mass, and metallicity of hSDF0705.25 is fully consistent with the FMR defined in \citet{2010MNRAS.408.2115M,2011MNRAS.414.1263M}. In Figure \ref{fig_fmr}, we plot the metallicity against the projection parameter $\mu_{0.32}=\log(M_*)-0.32\log(SFR)$ defined in \citet{2010MNRAS.408.2115M}. {The SFR is not corrected for host extinction. For $A_V=0.4$~mag, the correction would amount to a difference of 0.04 dex in $\mu_{0.32}$.}
{The metallicity in Figure \ref{fig_fmr} is derived from the [\ion{N}{ii}]/H$\alpha$ ratio and using the M08-calibration.}
% to the KD02-calibration using \cite{2008ApJ...681.1183K}. At metallicities above 8.4 the KD02-calibration is consistent with the calibration of \citet{2008A&A...488..463M} used in \citet{2010MNRAS.408.2115M,2011MNRAS.414.1263M}.
The hSDF0705.25 and the host of SN Primo at $z=1.55$ \citep{Frederiksen12} are the two SNe~Ia above $z>1$ for which a measurement of gas-phase metallicity has been obtained. For comparison we also plot the sample of gamma-ray burst (GRB) host galaxies {at $0.01<z<1$ from} \citet{2011MNRAS.414.1263M} and a sample of lensed galaxies presented in \citet[$1.5<z<3.5$]{2011MNRAS.413..643R} and \citet[$1.5<z<3.5$]{2012MNRAS.427.1953C}. {The fact that high-redshift SN Ia host galaxies follow the FMR relation suggests that they follow the same evolutionary path as regular field galaxies.}

{The connection between the Hubble residuals of SNe Ia and their hosts has been well documented recently \citep{2010ApJ...722..566L,2010MNRAS.406..782S,2011ApJ...737..102S,2013arXiv1304.4720C,2014arXiv1401.3352K}. The correlation between the Hubble residuals of SNe Ia and the stellar masses of their hosts is proposed to be proxy for a more fundamental connection between metallicities of the SNe and the metallicities of their hosts \citep{2009Natur.460..869K,2003ApJ...590L..83T}. A study by \citet{2013ApJ...764..191H} showed that the scatter in the Hubble residual decreased when using the the $\mu_\alpha$ parameter of the FMR (i.e., metallicity) to correct for the host correlation. This supports the argument that metallicity is the underlying driver of the correlation between the SN host stellar mass and the Hubble residual.}

\section{CONCLUSIONS}
Our derived spectroscopic redshift, $z=1.54563\pm0.00027$, is in full agreement with the G11 photometric redshift of $z=1.552\pm0.018$ (Figure \ref{fig_sed}). We also exclude AGN activity as the source of the emission-line flux.
From the flux of the H$\alpha$ line we derive an observed SFR and from the emission line ratios we constrain the host-galaxy metallicity.

The SN~Ia rates at high redshift are dominated by small-number statistics. The three $z>1.4$ SNe Ia found with the {\it Hubble Space Telescope (HST)} in the {\it HST}/GOODS survey \citep{dahlen2008} belong to host galaxies with measured spectroscopic redshifts and no AGN activity. On the other hand, the larger Subaru/SDF sample includes ten SNe Ia at $z>1.5$, but their classification as SNe Ia relies on photometric redshifts which, at high redshifts, might be systematically offset. Two ongoing {\it HST} Multi-Cycle Treasury programs, the Cluster Lensing and Supernova survey with Hubble \citep{Postman2012} and the Cosmic Assembly Near-IR Deep Extragalactic Legacy Survey \citep{Grogin2011,Koekemoer2011} and the upcoming Frontiers Fields will find new SNe out to $z\approx2.5$ \citep{2012ApJ...746....5R,jones12,2014ApJ...783...28G}, but their samples will still be small and will suffer from the same classification challenges faced by the GOODS and SDF surveys.
{For example, in the {\it HST}/CLASH survey, 2 of 4 $z>1.2$ SN Ia with light curves in multiple filters have spectroscopic redshifts \citep{2014ApJ...783...28G}
It is thus important to test the robustness of the SDF SN Ia rate measurements by spectroscopically measuring the redshifts of the SN host galaxies, and ascertaining whether the SN sample was contaminated by unclassified AGNs.
The confirmation of the photometric redshift of hSDF0705.25, and its classification as a star-forming, non-AGN-hosting galaxy, represents the first step in this endeavour.
hSDF0705.25 is the brightest of the high-redshift SDF SN Ia host galaxies, so to investigate other high-redshift SDF SNe hosts will require longer exposure times on ground based near-IR spectrographs (like X-shooter).
%The early-type galaxies at these redshifts can not be targeted from the ground and therefore require grism observations with {\it HST}.
{The study of early-type galaxies at these redshifts is currently beyond the capabilities of single object, ground based IR spectrographs, and will require observations from the future generation of ground based and space based instruments.}

\begin{acknowledgements}
{We thank the anonymous referee for valuable comments that improved this manuscript.} We also thank Martin Sparre for providing his X-shooter meta-pipeline, which simplified the reduction of the X-shooter spectra significantly.
The Dark Cosmology Centre is funded by the Danish National Research Foundation.
DM and OG acknowledge support by a grant from the Israel Science Foundation.
This research has made use of the NASA/IPAC Extragalactic Database (NED) which is operated by the Jet Propulsion Laboratory, California Institute of Technology, under contract with the National Aeronautics and Space Administration.
\end{acknowledgements}

%-------------------------------------------------------------------

\bibliography{graur}{}
\bibliographystyle{aa}

\end{document}